\documentclass[12pt,a4paper]{article}
\usepackage{amsmath,amssymb,cite}
\title {\textbf
{Hidden connection between\\general relativity and Finsler geometry}}

\author {Mehrdad Panahi\thanks{Email:  panahi@aeoi.org.ir}\\
\textit{Centre for Theoretical Physics and Mathematics,} \\
\textit{AEOI, P.O. Box 11365--8486, Tehran, Iran}}
\date{}
\begin{document}
\maketitle
\begin{abstract}
	Modern formulation of Finsler geometry of a manifold $M$
utilizes the equivalence between this geometry and the Riemannian
geometry of $VTM,$ the vertical bundle over the tangent bundle of $M,$
treating $TM$ as the base space.
	We argue that this approach is unsatisfactory when there is an
indefinite metric on $M$ because the corresponding Finsler fundamental
function would not be differentiable over $TM$ (even without its zero
section) and therefore $TM$ cannot serve as the base space.
	We then make the simple observation that for any
differentiable Lorentzian metric on a smooth space-time, the
corresponding Finsler fundamental function is differentiable exactly
on a proper subbundle of $TM$.  This subbundle is then used, in place
of $TM,$ to provide a satisfactory basis for modern Finsler geometry
of manifolds with Lorentzian metrics.
	Interestingly, this Finslerian property of Lorentzian metrics
does not seem to exist for general \textit{indefinite\/} Finsler
metrics and thus, Lorentzian metrics appear to be of special relevance
to Finsler geometry.
	We note that, in contrast to the traditional formulation of
Finsler geometry, having a Lorentzian metric in the modern setting
does not imply reduction to pseudo-Riemannian geometry because metric
and connection are entirely disentangled in the modern formulation and
there is a new indispensable \textit{non-linear connection\/},
necessary for construction of Finsler tensor bundles.
	It is concluded that general relativity\textemdash without any
modification\textemdash has a close bearing on Finsler geometry and a
modern Finsler formulation of the theory is an appealing idea.
	Furthermore, in any such attempt, the metric should probably
be left unchanged (not generalized) or the newly discovered
property, which provides a satisfactory basis for the corresponding
Finsler geometry, would be lost.
\end{abstract}

\newpage
	
	\section{Introduction}
	Finsler geometry is widely considered as the most natural
generalization of Riemannian geometry or even closer (just a ``more
developed'' form \cite{Chern1}).
	On the other hand, general relativity is the first and by
far the most important physical application of (pseudo-) Riemannian
geometry.
	It is therefore natural to seek for a viable Finsler
formulation of general relativity.
	Indeed, there is a long and extensive history of research in
this area and the field has been steadily growing to this date.
See, \textit{e.g.\/}, \cite{set1,Bek,set2,Bogos,set3} and the
references therein.

	Finsler geometry has originated from two simple innovations
in Riemannian geometry, namely\footnote[1]{For a lucid exposition to
modern Finsler geometry, see \cite{Bejancu} and for a recent account
of the classical treatment, consult \cite{Asanov}.}:

\begin{enumerate}
\item[(a)]
Supplementation of the position parameter in geometric quantities
with a new independent vector variable.  Here, this is given the name
``Finsler parameter''.
\item[(b)]
Use of a norm, here called ``Finsler fundamental function''
(a scalar distance function of position and the Finsler parameter)
in order to implement the first technique.
\end{enumerate}
	As we shall see, a Finsler fundamental function is equivalent
to a metric tensor (see equations (\ref{g}) and (\ref{F2})) and some
authors use the term metric to mean a fundamental function, however,
for more clarity, the term metric is used in this article only to
mean a metric tensor.

	The range of Finsler parameter is usually assumed to be all
non-zero tangent vectors and skipped over quickly, however, the
subject merits more attention, particularly, in the case of metrics
with indefinite signatures.
	Finsler parameter can be present in intrinsic geometric
quantities such as connections and curvatures, which in a
differential geometric context, all need to be differentiable,
albeit, not infinitely.
	It is therefore necessary that this parameter takes
\textit{only\/} values for which all such quantities are well-behaved.
	The most natural and practical way to determine the range of
Finsler parameter is evidently through Finsler fundamental function.
	Domain of differentiability of this function seems the best
(and the only available) candidate for the purpose.   However, life
is not that simple.
	As we shall see, the range of Finsler parameter has
to be a fibre bundle in order to obtain a vertical bundle, absolutely
necessary in the modern formulation.
	When the metric is positive definite this requirement is
easily satisfied because the corresponding fundamental function is
differentiable for all non-zero tangent vectors, which form a fibre
bundle.
	This is the prevailing situation in most studies of Finsler
geometry and its applications.
	However, for an indefinite metric, domain of differentiability
of the fundamental function is more restricted and it is not clear if
it forms a fibre bundle in general.

	\section{A brief review of modern Finsler geometry}
	In order to have a closer look at modern formulation of 
Finsler geometry, let $M$ be a manifold and $TM$ its tangent bundle.
      Also let $x$ represent a point of $M$ with coordinates
$\{x^{\mu} \},$  $(x,y)$ a point of $TM,$ and $\{ y^{\mu}\}$ 
coordinates of the tangent vector $y$ with respect to the natural 
basis $\{\partial / \partial x^{\mu} \}.$
	Einstein summation convention is used throughout.
	Being merely a tangent vector to $M,$ Finsler parameter has 
classically no proper geometrical basis to work with.
	This is clearly illustrated by the classical ``Finsler vector 
field'' $X^{\mu} (x,y)(\partial / \partial x^{\mu})_{M}$ on $M,$ which
is not in fact a true vector field on $M$(an assignment of at most 
\textit{one\/} vector to each point $x$).
	The most direct and natural attempt to accommodate Finsler 
parameter geometrically has been to consider $TM$ as the base space 
and Finsler vector fields as sections of $\pi ^{*} TM$ (the pull back 
of $TM$ to itself by its own projection $\pi$).
	Although some authors may be happy with such an improvement, 
this is still unsatisfactory and cumbersome to work with because 
sections of $\pi^{*}TM$ are not \textit{tangent\/} vector fields to 
$TM$\footnote{Fibres of $\pi^{*}TM$ are spanned by basis vectors 
$(\partial / \partial x^{\mu})_{x},$ which are tangent only to $M.$
These vectors should not be confused with similar looking objects 
$(\partial / \partial x^{\mu})_{(x,y)},$ which transform differently 
under a coordinate transformation of $M.$ See, \textit{e.g.\/},
\cite[p.~11]{Bejancu}.}.
	And similarly, sections of the dual bundle cannot be 
considered as differential forms on $TM$\footnote{Differential 
1-forms on $TM$ are sections of only $T^{*}TM.$}.
	The natural isomorphism between $\pi^{*}TM$ and $VTM$ (the
vertical bundle over $TM$) \cite[p.~18]{Poor} helps to remedy the
situation by mapping sections of $\pi^{*}TM$ to sections $X^{\mu}
(x,y)(\partial / \partial y^{\mu})_{TM}$ of $VTM,$ which are
\textit{tangent\/} vector fields to $TM.$
	Riemannian geometry of $VTM$ would thus provide a lucid and
satisfactory framework for Finsler geometry of $M.$
	This is what is meant by modern formulation of Finsler
geometry in this article.
	There is a minor variation of this formulation in which the 
zero section of $TM$ is removed \cite{Abate}.
 
	The basic limitation in this modern $VTM$--formulation, which 
we wish to point out, is due to the implicit assumption that:  
\textit{All (non-zero) tangent vectors of $M$ are admissible values 
for the Finsler parameter.\/}
	As explained in the introduction, this assumption is justified
when we have a positive definite metric or none at all.
	However, for any space with an indefinite metric, the 
corresponding Finsler fundamental function is not differentiable 
over $TM$ or $TM \backslash \{ \text{zero section}\}$ and hence none
of these would be a suitable basis for formulating the corresponding 
Finsler geometry.
	Our new simple result (see next section) offers exactly the 
further improvement that we need for the important case of Lorentzian 
metrics.  

	\section{A new result}
	We need only a few basic relations in Finsler geometry. 
	To collect these, let $N$ be some open submanifold of tangent 
bundle $TM.$     
      A Finsler fundamental function is defined as a map 
$F \colon\ N \to \mathbb{R},$ satisfying a varying set of conditions.
Naturally, first-degree homogeneity in $y$ is nearly always among
these conditions,
		\begin{equation}  
F(x,ky) = k F(x,y),  \;  \forall k>0,  \; \forall (x,y){\in} N,
		\label{homo}
		\end{equation}
where, it is implicitly assumed that if $ (x,y){\in} N$ then so is
$ (x,ky)\ \forall   k>0.$
	Some authors restrict choice of $N$ to only $TM,$ however,
important classes of Finsler spaces would be lost by this restriction
\cite[p.~13]{Bejancu}.
	Applying Euler theorem on homogeneous functions to $F$ yields:
		\begin{equation}
F^{2}= \frac{1}{ 2}\   \frac{\partial ^{2} F^{2}}
{\partial y^{\mu} \partial y^{\nu}}\   y^{\mu} y^{\nu}.
		\label{Euler}
		\end{equation}
	Finsler metric tensor is classically defined by:
		\begin{equation}
	G_{\mu\nu}(x,y)  = \frac{1}{ 2}\
	\frac{\partial^{2} F^{2}} {\partial y^{\mu} \partial y^{\nu}},
		\label{g}
		\end{equation}
where, the $y$-Hessian of $F^{2}$ is assumed to be of maximal rank
and hence of fixed \textit{signature\/}.
	$G_{\mu\nu}$ may be definable only on a subset of $N$ because
$F$ may be not differentiable on the whole of $N.$
	Combining equations (\ref{Euler}) and (\ref{g}) yields the
important relation:
		\begin{equation}
F^{2}(x,y) = G_{\mu\nu}(x,y)\    y^{\mu} y^{\nu}.
		\label{F2}
		\end{equation}
	Alternative to the classical approach, given any arbitrary
zero-degree $y$-homogeneous Finsler metric tensor, we can consider
equation (\ref{F2}) as the \textit{definition\/}  of the Finsler
fundamental function \textit{corresponding\/} to the given metric.
	Property (\ref{Euler}) ensures that equations (\ref{g}) and
(\ref{F2}) are not only consistent, but equivalent, given only that 
the metric tensor in equation (\ref{F2}) is zero-degree 
$y$-homogeneous.
	Needless to say, any (pseudo-) Riemannian metric is also a
$y$-homogeneous Finsler metric, albeit a special one.
	In what follows, ``differentiable'' shall mean
differentiable of class $C^{k}$ with $k$ as large as necessary, and
``Lorentzian metric'', a pseudo-Riemannian metric of signature
$(+$$-$$-$$-).$
	We can now state an intriguing result which has a simple 
proof:\\ \\
	\noindent
	\textbf{Proposition  1   }
Given any differentiable Lorentzian metric on a smooth space-time, the
corresponding Finsler fundamental function is differentiable
\textit{exactly on a fibre bundle\/} over the space-time.\\ \\
	\noindent
	\textbf{Proof     }
	Let $M$ be a $C^{\infty}$ space-time manifold with a
differentiable Lorentzian metric $g_{\mu\nu}(x)$ and assume that the 
corresponding Finsler fundamental function
$F(x,y):= (g_{\mu\nu}\ y^{\mu} y^{\nu})^{1/2}$ is defined over the
largest possible domain
$N:=\{ (x,y) {\in}TM \mid g_{\mu\nu}\ y^{\mu} y^{\nu} \geq 0 \}.$
	Clearly, $N$ has a boundary in $TM$ given by
$\{(x,y) {\in} TM \mid  g_{\mu\nu}\ y^{\mu} y^{\nu}  = 0 \}$ at which
$F$ cannot, by definition, be differentiable\footnote{Clearly, at no
boundary point, can all directional derivatives of a function exist,
see, \textit{e.g.\/}, \cite[p.~5]{Warner} or \cite[p.~349]{Bartle}.}.
	Elsewhere in $N,$ $F$ can easily be seen to be differentiable.
	Hence, $F$ is differentiable  \textit{exactly on:\/}
		\begin{equation}
	LM:=\{ (x,y) \in TM \mid g_{\mu\nu}\ y^{\mu} y^{\nu}  > 0 \}.
		\label{LM}
		\end{equation}
	We see that $LM$ is simply made of all timelike vectors,
yet surprisingly, no proof or statement to the effect that $LM$ is in
fact a fibre bundle, is found in the literature.
	Here is, therefore, a detailed proof of that, wherein, the
definition for a fibre bundle is taken from \cite{Poor} and followed
closely:

\begin{enumerate}
\item[(i)]
	Three $C^{\infty}$ manifolds are needed to start with.
	The base manifold is already given and the total space $LM$ is
an open subset of the $C^{\infty}$   manifold $TM$ and hence, a
$C^{\infty}$ manifold \cite[p.~7]{Warner}.
	As for the standard fibre, let $V$ be a real four-dimensional
vector space with an inner product $\eta$ of signature $(+$$-$$-$$-)$
and define $L$ to be the open subset
$\{v {\in} V \mid \eta(v,v) >0 \}.$  Being a real vector space, $V$ is
also a $C^{\infty}$ manifold \cite[p.~7]{Warner} and hence so is $L,$
which would be our standard fibre for $LM.$

\item[(ii)]
	Given the projection map $\pi \colon\ TM \to M,$ its
restriction $\chi :=\pi |_{_{LM}}$ serves as the projection for $LM.$

\item[(iii)]
	Let $\mathcal{C}$ be an open covering of $M$ such that there
is a complete set of orthonormal basis vector fields defined on each
$U \in \mathcal{C}.$  Denote one such frame on  $U \in \mathcal{C}$
by $\{e_{a}\}$ and the corresponding co-frame by $\{e^a\}$:
$e^{a}(e_{b})={\delta^{a}}_{b},\ a,b=0,1,2,3.$
	Define $\varphi \colon\ \chi^{-1} U \to L$ by
$\varphi (x,y)=e^{a}(y)f_{a},$
where, $\{f_{a}\}$ is an orthonormal basis for $V$ with
$\eta(f_{a},f_{b})=g(e_{a},e_{b}).$
For any $x {\in} U,$ the map $\varphi |_{x} \colon\ \chi^{-1}(x)\to L$
is clearly a diffeomorphism, and hence, so is
$ (\chi,\varphi) \colon\ \chi^{-1}U  \to U \times L.$
	Local triviality of $LM$ is thus established.
	\hspace*{\fill} $\Box$\\
\end{enumerate}
	\section{Modern Finsler geometry for Lorentzian metrics}
	To distinguish Finsler parameter from an arbitrary tangent
vector, let us denote it from now on by $z$ rather than $y.$
	Given any Lorentzian metric, the crucial step we now take is 
to let bundle $LM,$ defined by equation (\ref{LM}), be our ``Finsler 
base space'': the space of admissible values of $z$ or more precisely 
$(x,z).$  
	The justification is that the corresponding fundamental
function is differentiable (and non-zero) \textit{exactly on\/} $LM.$
	More importantly, being a fibre bundle, $LM$ is endowed with a
natural vertical bundle $VLM$ \cite{Poor}, which has all the crucial 
properties of $VTM$:\\ \\
	\noindent
	\textbf{Proposition  2  }
	Fibres of $VLM,$ $VTM$ and $TM$ are isomorphic as vector
spaces and have the same coordinate transformations under changes of
coordinates on $M.$\\ \\
	\noindent
	\textbf{Proof   }	
	$VTM$ and $TM$ already have isomorphic fibres with the same 
coordinate transformations \cite[p.~19]{Bejancu}.
	To prove that $VLM$ is also in this category, it suffices to
note that fibres of $VLM$ and $VTM$ are tangent spaces to fibres of
$LM$ and $TM$ respectively, and that, fibres of $LM$ are open subsets
of fibres of $TM.$
	Therefore, fibres of $VLM$ are in fact also fibres of $VTM$
and have the same properties. \hspace*{\fill}     $\Box$\\

	Through this simple modification, Finsler parameter has been 
effectively raised to the status of a coordinate parameter and 
sections $X^{\mu} (x,z)(\partial / \partial y^{\mu})_{TM}$
are traded for the more natural ones
$X^{\mu} (x,z)(\partial / \partial z^{\mu})_{LM}.$
	It is easy to verify that further constructions, such as
non-linear connection, Finsler connection, etc., can all be obtained
for $VLM$ in a straightforward manner.
	See, \textit{e.g.\/},  \cite[p.~109]{Bejancu} for a general
setup.

	An important feature of modern Finsler geometry is that the 
treatments of connection and metric are conveniently decoupled and
in such a way that the geometry can \textit{never\/} be reduced to
Riemannian geometry of the original manifold even when the metric is
Riemannian:  \label{explanation}
	There is a new indispensable \textit{non-linear connection\/},
which is vital for the construction of \textit{dual\/} spaces to
fibres of the above vertical bundles \cite[p.~111]{Bejancu} and hence
also essential in the construction of the necessary vertical tensor
bundles.
	This feature of modern Finsler geometry is in direct contrast
to the widely used classical formulation, where every thing hinges on
a Finsler metric tensor such that reduction of this to a Riemannian
metric implies reduction to Riemannian geometry 
\cite{Chern1,Bek,Bogos}.

	It is interesting to note that, the above approach yields a
satisfactory basis for modern Finsler geometry \textit{only\/} for
$z$-independent metrics.
	While, for any $z$-dependent \textit{indefinite\/} metric,
a corresponding fundamental function exists, such a function would not
in general be differentiable on the whole of $TM,$ nor on any other
known fibre bundle.
	For example, assuming that $F$ is differentiable for
$  G_{\mu\nu}(x,z)\ z^{\mu} z^{\nu} >0,$ the spaces
$\{ (x,z) {\in} T_{x}M \mid G_{\mu\nu}(x,z)\ z^{\mu} z^{\nu} >0 \}$ 
would depend so non-trivially on $z$ that it is hard to imagine how 
they can form a fibre bundle in general.
	Trying to remedy this situation by resorting to the classical
strategy of starting with a general \textit{indefinite\/} $F$ instead,
would not help because we still need a Finsler base space, in the form
of a fibre bundle over which $F$ is differentiable, in order to obtain
the necessary vertical bundles.
	Consequently, \textit{an appropriate framework for modern 
Finsler geometry of general indefinite metrics has yet to be found.\/}

	On the other hand, for Lorentzian metrics, the simple proof of
 proposition~1 allows the conditions of global space-time smoothness 
and metric differentiability to be reduced to local ones.
	Thus, the above approach based on $VLM,$ is generalizable to 
even \textit{local\/} Lorentzian metrics.
	The generalized form of proposition~1 would be:\\ \\
	\noindent
	\textbf{Proposition  3   }
	Given a Lorentzian metric, differentiable over a smooth open 
subset $U$ of space-time, the corresponding Finsler fundamental 
function, restricted to $U,$ is differentiable exactly on a fibre 
bundle over $U.$
	\hspace*{\fill}  $\Box$
	\section{Conclusions}
	In conclusion, we see that general relativity\textemdash
without any modification\textemdash has a close bearing on Finsler
geometry.
	Propositions~1-3 provide some ``mathematical'' evidence to 
support this conclusion.
	The generality and inherent simplicity of these propositions
indicate that the connection between general relativity and Finsler
geometry is not artificial or lightly dispensable.
	Accordingly, searching for a viable modern Finsler formulation
of general relativity would now seem more appealing and hopeful than
before.
	It seems as though general relativity has some special
built-in provisions for Finsler geometry and it would be interesting
to see if there are any further clues in this avenue yet to be
discovered.

	The second conclusion is that, in any viable Finsler
formulation of general relativity, the Lorentzian metric should
probably not be \textit{generalized\/} to a $z$-dependent one or the
newly discovered connection, which provides a satisfactory basis for
the corresponding Finsler geometry, would be lost.
	Naturally, this point can greatly simplify the search for
a viable Finsler formulation of general relativity, albeit a great
deal of work still remains.
	\section*{Acknowledgments}
	The author is grateful to Mohammad Mehrafarin, Rahim
Zaare-Nahandi and Vahid Aali for several helpful discussions and to 
Mehran Panahi for providing some of the references.

\end{document}